# A Note on Why Geographically Weighted Regression Overcomes Multidimensional-Kernel-Based Varying-Coefficient Model

Zihao Yuan[1]

It is widely known that geographically weighted regression(**GWR**) is essentially same as varying-coefficient model. In the former research about varying-coefficient model, scholars tend to use multidimensional-kernel-based locally weighted estimation(**MLWE**) so that information of both distance and direction is considered. However, when we construct the local weight matrix of geographically weighted estimation, distance among the locations in the neighbor is the only factor controlling the value of entries of weight matrix. In other word, estimation of **GWR** is distance-kernel-based. Thus, in this paper, under stationary and limited dependent data with multidimensional subscripts, we analyze the local mean squared properties of without any assumption of the form of coefficient functions and compare it with **MLWE**. According to the theoretical and simulation results, geographically-weighted locally linear estimation(**GWLE**) is asymptotically more efficient than **MLWE**. Furthermore, a relationship between optimal bandwith selection and design of scale parameters is also obtained.

## 1.Introduction

Spatial heterogeneity is always a key point in the research of spatial analysis and statistics. One of the most popular method we use to capture heterogeneity is spatially varying-coefficient model.

[1] Graduated Student(Ms), School of Data Science, Zhejiang University of Finance and Economics, Hangzhou, China

Compared with traditional varying-coefficient model, here we use variable of coefficient function, denoted by a d-dimensional real vector **U**, to represent locations of each individual observed. Under most of circumstances, d is equal to 2 or 3, since we tend to assume heterogeneity is caused by different geographical positions. In 1996, Fortheringham and his colleagues proposed geographically weighted regression(hereafter **GWR**), which is of great appeal for the analysis of spatial data. In **GWR** the spatial heterogeneity of a regression relationship is explored by locally calibrating a spatially varying coefficient model of the form

$$y_i = \sum_{k=1}^{p} \beta_k(u_i, v_i) x_{ik} + \varepsilon_i, \quad (1\text{-}1)$$

where $(u_i, v_i)$ denotes the longitude and latitude of the ith location, i=1,2,...,n. $\beta(u,v)$ is an unknown smoothing function. The model allows a geographically varying intercept by setting $x_{i1} \equiv 1$ (i=1,2,...,n).

Except for the great first-step work by Brundson et al(1996; 1998; 1999a) and Fortheringham et al (1997a; 1997b). **GWR** has been well studied from multiple directions. Fortheringham et al(1998) compared **GWR** with expansion while Brundson et al(1999b) with random coefficient model. Leung et al(2000a; 2000b) discussed the hypothesis tests for spatial nonstationarity of the coefficients among the residuals based on **GWR** method. Páez et al(2002a; 2002b) proposed a maximum likelihood estimator for **GWR** with error-variance heterogeneity. Fortheringham et al(2002) systematically made a nice review of the basic theory and statistical inference problems of **GWR** and proposed some possible extensions of **GWR**, like logit/probit **GWR**. In 2005, Wheeler and Tiefelsdorf fully discussed the multicollinearity and correlation of this model.

However, methodologically speaking, traditional **GWR** estimation is essentially same as Nadayara-Watson(hereafter **NW**) estimation, where we use a constant to locally fit(for example, see Härdle, 1990; Wand and Jones, 1995). It is well known that **NW** estimation generally leads to more obvious bias, particularly at the boundary points. Fan and Gijbels(1996) illustrated this phenomenon by noting that near the boundary "less observations contribute in computing the estimators". Meanwhile, the boundary-effect problem for **GWR** has also been addressed by Brudson(1996) and Fortheringham et al(1997a). More recently, Wang et al(2008) combined locally linear fitting and **GWR** methods together and proved local linear estimation is unbiased when coefficient function is a linear function. But due to our limited resource of literature, there has been few work discussing the locally asymptotic properties of geographically-weighted locally linear estimation(**GWLE**) with only some assumptions of smoothing conditions of coefficient function.

On the other hand, as we can tell from (1-1), traditional **GWR** method only considers 2-dimensional locations. Huang et al(2010) took time position into consideration and proposed geographically and temporally weighted regression(hereafter **GTWR**). A significant of characteristic of **GWR** or **GTWR** is, when we calculate the weight matrix of each location, a metric function is frequently used to measure the distance between any two locations. Since the distance function is actually a mapping from $\mathbf{R}^d$ to $\mathbf{R}^+$, we only need to consider 1-dimensional bandwith and kernel function. Specifically, for any two d-dimensional locations, $\mathbf{U}_1$ and $\mathbf{U}_2$, the distance function, denoted by $d(\cdot)$, frequently has the following form,

$$d(\mathbf{U}_1, \mathbf{U}_2) = \sqrt{(\Delta \mathbf{U})\Lambda(\Delta \mathbf{U}^T)}, \qquad (1\text{-}2)$$

where $\Lambda = \text{diag}(a_1, a_2, ..., a_d)$ is the scale parameter vector, $\Delta \mathbf{U} = \mathbf{U}_1 - \mathbf{U}_2$. Obviously, the non-zero entries of $\Lambda$ are used to balance the contribution of distance in each dimension. However, we have not found many previous work in comparing this distance-kernel-based idea with traditional **MLWE** theoretically. Additionally, research about how the scale parameters affect the conditional bias and variance is rare as well. Therefore, in this paper, based on the distance function with form (1-2), we compare **GWLE** with **MLWE** by investigating the leading terms of conditional bias and variance of locally linear estimation, which could also reflect the effect from scale parameters to prediction accuracy. In section 2, basic notations, assumptions and construction of estimator is presented. In section 3, we demonstrate the preliminary lemmas, main results and proof of the major theorems. These results would theoretically show that, based on the assumptions in section 2, **GWLE** ought to be asymptotically more efficient than **MLWE**. Besides, by using minimizing integrated mean-square error(**IMSE**) as the goal function, a relationship between optimal bandwith selection and design of scale parameters is also obtained. Section 4 will be the conclusion part. At last, technique proof of lemmas will be shown in appendix.

Meanwhile, to expand the adaptability of our discussion, we assume the sample sequence can be expressed as $\{(\mathbf{X_i}, \mathbf{U_i}, y_i) | \mathbf{i} \in Z^{M+}\}$ and is limited dependent over all directions. Most of the literature about **GWR** or **GTWR** is based on a sample with 1-dimensional subscript, $\{(\mathbf{X}_i, \mathbf{U}_i, Y_i) | i = 1,2,...,n\}$. However, sometimes panel data and multidimensional panel data are also widely-seen in our research. Actually, any sample from one of these kinds of data can be considered as a sample from random field with rectangular position which can be written as $\{(\mathbf{X_i}, \mathbf{U_i}, Y_i) | \mathbf{i} \in Z^{M+}\}$, where $Z^{M+}$ is the set of M-dimensional positive integers. M itself is also a positive integer indicating the quantity of axis of coordinates or directions.

## 2. Estimation

Firstly, considering the convenience of statement, we use the following notations. About the derivatives of multivariate function and matrix, we use the following rules of expression, for any d-dimensional vector $\mathbf{u} = (u_1,..u_d)$, positive integer d and p, real and measurable mapping $f: \mathbf{R}^d \to \mathbf{R}$ and $\Omega: \mathbf{R}^d \to \mathbf{R}^{p \times p}$, by setting $\mathbf{u} = \mathbf{u}_0$

$f'(\mathbf{u}_0) = (f_1'(\mathbf{u}_0), f_2'(\mathbf{u}_0),..., f_d'(\mathbf{u}_0))^T$, $\Omega'(\mathbf{u}_0) = (\Omega_1^T(\mathbf{u}_0),...,\Omega_d^T(\mathbf{u}_0))^T$ where the ith entry(block) denote $f_i'(\mathbf{u}_0) = \frac{\partial f}{\partial u_i}\big|_{\mathbf{u}=\mathbf{u}_0}$, $\Omega_i'(\mathbf{u}_0) = \frac{\partial \Omega}{\partial u_i}\big|_{\mathbf{u}=\mathbf{u}_0}$. Please note $\Omega_i'(\mathbf{u}_0)$ is a $p \times p$ matrix. Meanwhile, since the subscript i is a vector, for any kth direction, we set the sample size as $N_k$. Then the total sample size should be $\prod_{k=1}^{M} N_k := \tilde{N}$.

Secondly, we present the assumptions which are the foundation of the incoming theoretical discussion.

**A1.** Sample sequence $\{(X_\mathbf{i}, \mathbf{U}_\mathbf{i}) | \mathbf{i} \in Z^{M+}\}$ is stationary and satisfies the following dependent structure, for any given $\mathbf{i}$ and a positive integer m, if $\|\mathbf{j} - \mathbf{i}\|_\infty > m$, the two random vectors $(X_\mathbf{i}, \mathbf{U}_\mathbf{i})$ and $(X_\mathbf{j}, \mathbf{U}_\mathbf{j})$ are independent from each other, where $\|\ \|_\infty$ denotes infinite norm. Additionally, we assume there also exists the following inequality,

$$\sup_{\mathbf{i}} \sup_{\mathbf{j}:\|\mathbf{j}-\mathbf{i}\|_\infty \leq m} \left| \frac{Cov(\phi(X_\mathbf{i}, \mathbf{U}_\mathbf{i}), \phi(X_\mathbf{j}, \mathbf{U}_\mathbf{j}))}{\sqrt{Var(\phi(X_\mathbf{i}, \mathbf{U}_\mathbf{i}))}\sqrt{Var(\phi(X_\mathbf{j}, \mathbf{U}_\mathbf{j}))}} \right| := \rho(m) < +\infty,$$

where $\phi(.)$ is any measurable function.

**A2.** Coefficient function $\beta_k(.)$ has continuous second partial derivatives.

**A3.** Marginal density function of location $\mathbf{U}$, $f(\mathbf{u})$, has continuous partial derivatives and $\inf\{f(\mathbf{u})\} > 0$.

**A4.** $\Omega_l(\mathbf{u})$, $l = 0,1,2$, and $E((X_{i_1})^{j_1}(X_{i_2})^{j_2} | \mathbf{U} = \mathbf{u})$ has continuous partial derivatives, where $E(X \mid \mathbf{U} = \mathbf{u}) = \Gamma(\mathbf{u})$, $E(X^\mathrm{T} X \mid \mathbf{U} = \mathbf{u}) = \Omega(\mathbf{u})$, $E(X^\mathrm{T} X * X^\mathrm{T} X \mid \mathbf{U} = \mathbf{u}) = \Omega_2(\mathbf{u})$,

$1 < i_1, i_2 < p$, $j_1 + j_2 = 4$.

**A5**. The kernel function $K_h(\cdot/h)$ we choose, satisfies the following conditions:

(1) $K(v)$ can be rewritten as $\phi(v^2)$ and $\phi(\sum_{i=1}^n v_i) = \prod_{i=1}^n \phi(v_i)$.

(2) $K(v)$ is symmetrical and bounded density function with compact support set.

**A6**. As $\widetilde{N} \to +\infty$, we have $h \to 0$ and $\widetilde{N} h^{d-1} \to 0$.

**Remark 1**. For A1, infinite norm here means that, over each direction, range of dependence can not be larger than m. More explicitly, for any given i, based on infinite norm, number of js dependent on i is larger than it is based on euclidean norm or 1-norm. When M=2(panel data), we can illustrate this point as the following Figure a, where m is the length of dependence; blue denotes the boundary of dependence based on 1-norm, while red based on euclidean norm and black based on infinite norm.

**Remark 2**. The existence of A5 indicates that most of the kernel densities from exponential family satisfy this assumption, particularly Gaussian kernel function. The first assumed condition of A5(1) is obvious since when we run **GWR** algorithm, we frequently compute the weight between any two locations in this way. Meanwhile, the kernel function in this paper has this form, $K_h(\cdot/h) = K(\cdot/h)/h$. Although the form $K(\cdot/h)$ is also widely used(for example, see Wang(2008)), the differences between the two forms will not cause different results obtained in this paper. This point can be checked based on the procedure of our proof.

**Remark 3**. As for $\widetilde{N} \to +\infty$, we assume $\min_{1 \leq k \leq M} n_k \to +\infty$ and $(n_k / n_l) < +\infty$, $k \neq l$.

Finally, we construct our estimator. The model we discuss in this paper can be expressed as follow,

$$y_i = X_i \beta(\mathbf{U_i}) + \sqrt{\sigma}(\mathbf{U_i})\varepsilon_i, \qquad (2\text{-}1)$$

where $E(\varepsilon \mid X, \mathbf{U}) = 0$, $Var(\varepsilon) = 1$ and $\{\varepsilon_i\}$ is an identically distribution and independent sequence. Apparently, according to (1-1) and A1, the mean regression function is $E(y \mid X, \mathbf{U} = \mathbf{u}) = X\beta(\mathbf{u})$ and variance regression function is $Var(y \mid X, \mathbf{U} = \mathbf{u}) = \sigma(\mathbf{u})$. Hence, by applying the idea of locally linear estimation, for any given $\mathbf{u}_0 \in S(\mathbf{U})$ ($S(\mathbf{U})$ is the support set of vector $\mathbf{U}$), estimator can be constructed by minimizing the following goal function,

$$\arg\min \sum_{\mathbf{i} \in \mathbf{I}} K_h\left(\frac{d_0(\mathbf{u_i})}{h}\right)\left(y_i - \sum_{k=1}^{p} X_{ik}\left(\beta_k(\mathbf{u}_0) + \sum_{s=1}^{d} \beta_{ks}'(\mathbf{u}_0)(\mathbf{u}_{is} - \mathbf{u}_{0s})\right)\right)^2,$$

where $\mathbf{I}$ denotes the set of subscripts of sample and $d_0(\mathbf{u_i}) = d(\mathbf{u_i}, \mathbf{u}_0)$. By using some simple algebraic knowledge, our estimator can be constructed as follow,

$$\hat{\beta}(\mathbf{u}_0) = e_1^{\mathrm{T}}\left(\sum_{\mathbf{i} \in \mathbf{I}} K_h(\tfrac{d_0(\mathbf{u_i})}{h}) \widetilde{X}_\mathbf{i}^{\mathrm{T}} \widetilde{X}_\mathbf{i}\right)^{-1}\left(\sum_{\mathbf{i} \in \mathbf{I}} K_h(\tfrac{d_0(\mathbf{u_i})}{h}) \widetilde{X}_\mathbf{i}^{\mathrm{T}} y_\mathbf{i}\right), \qquad (2\text{-}2)$$

where $e_1$ is a $(d+1)p$-dimensional column vector whose first $p$ entries are 1 while the others are 0 and row vector $\widetilde{X}_\mathbf{i} = (X_\mathbf{i}, (\mathbf{u_i} - \mathbf{u}_0) \otimes X_\mathbf{i})$. Furthermore, in matrix form, we can rewrite this estimator as follow,

$$\hat{\beta}(\mathbf{u}_0) = e_1^{\mathrm{T}}\left(\widetilde{X}^{\mathrm{T}} W(\mathbf{u}_0) \widetilde{X}\right)^{-1}\left(\widetilde{X}^{\mathrm{T}} W(\mathbf{u}_0) y\right), \qquad (2\text{-}3)$$

where $\widetilde{X} = \begin{bmatrix} X_1 & (\mathbf{u_1} - \mathbf{u}_0) \otimes X_1 \\ \vdots & \vdots \\ X_n & (\mathbf{u_n} - \mathbf{u}_0) \otimes X_n \end{bmatrix}$, $\mathbf{1}$ is a $M$-dimensional subscript whose entries are all 1, $\mathbf{n} = (n_1, .. n_M)$, and $W(\mathbf{u}_0) = diag(K_h(\tfrac{d_0(\mathbf{u_1})}{h}), ..., K_h(\tfrac{d_0(\mathbf{u_1})}{h}))$. It is apparent that, according to the construction of **GWLE**, this is also essentially same as pool least square estimation in panel data model. Therefore, we personally call this kind of estimator as geographically weighted linear pool least square estimation.

## 3. Main Results

In this section, our major two tasks are: 1. obtaining the leading terms of conditional bias and variance of **GWLE**; 2. obtaining globally optimal design of scale parameters. In order to simplify the complexity of discussion, we here only consider the situation when $\mathbf{u}_0$ is an interior point. In other word, given a bandwith h and $\mathbf{u} = \mathbf{u}_0$, the d-dimensional ellipsoid $\{\mathbf{u}:(\mathbf{u}-\mathbf{u}_0)\Lambda(\mathbf{u}-\mathbf{u}_0)^T \leq h^2\} := Nei(\mathbf{u}_0)$ is a subset of the support set of vector U, S(U). Recall the core idea of **GWR** or **GTWR** is to borrow the data from neighbors. Owing to different neighbors for different locations, information borrowed would be varying. This mechanism significantly reflect the existence of heterogeneity. Furthermore, as the quantity of information borrowed tends to be infinitely large ($\widetilde{N} \to +\infty$), how do we ensure the consistency of estimator? A conceptually understanding is as follow, please note that the largest axial length of this ellipsoid is $(a_{max})^{-1}h^2$, where $a_{max} = \max\{a_1,...,a_d\}$. Since $h \to 0$ as $\widetilde{N} \to +\infty$, $Nei(\mathbf{u}_0)$ will converges $\{\mathbf{u}_0\}$. This property is the key point for the consistency of many kinds of locally weighted estimators which will be reflected in the proof of theorem 1. Meanwhile, we introduce the following lemmas as preliminary work.

**Lemma 1**. *For any given $s \in \{1,2,...,d\}$ and $\mathbf{u}_0 \in \{\mathbf{u}: Nei(\mathbf{u}) \subset S(\mathbf{U})\}$, under assumptions from A1 to A6, by setting $A^s_{\widetilde{N}\lambda} = \widetilde{N}^{-1}\sum_{\mathbf{i}\in\mathbf{I}} K_h(\frac{d_0(\mathbf{U_i})}{h})h^{-l}(\mathbf{U_{is}} - u_{0s})^\lambda X_\mathbf{i}^T X_\mathbf{i}$, $\lambda=0,1,2$ and l=d-1 or d+1, we have*

$$A^s_{\widetilde{N}\lambda} = (\det(\Lambda)a_s^\lambda)^{-1} h^{(\lambda+d)-(l+1)} \kappa_\lambda \Omega(\mathbf{u}_0) f(\mathbf{u}_0) \\ + (\det(\Lambda)a_s^{\lambda+1})^{-1} h^{\lambda+d-l} \kappa_{\lambda+1}\{\Omega'_s(\mathbf{u}_0)f(\mathbf{u}_0) + \Omega(\mathbf{u}_0)f'_s(\mathbf{u}_0)\} + o_p(1) \quad (3\text{-}1)$$

where $\kappa_\lambda = \int z^\lambda K(z)dz$, $\Omega'_s(\mathbf{u}_0) = \frac{\partial\Omega}{\partial u_s}\big|_{\mathbf{u}_0}$, $f'_s(\mathbf{u}_0) = \frac{\partial f}{\partial u_s}\big|_{\mathbf{u}_0}$.

**Lemma 2**. *With the same preconditions as lemma 1, while $\lambda=0,1$, we have the following results,*

set $B^s_{\widetilde{N}\lambda} = \widetilde{N}^{-1}\sum_{\mathbf{i}\in\mathbf{I}} K_h(\frac{d_0(\mathbf{U_i})}{h})h^{-l}(\mathbf{U_{is}} - u_{0s})^\lambda X_\mathbf{i}^T X_\mathbf{i} \Pi(\mathbf{U_i},\mathbf{u}_0)$, then

$$B^s_{\widetilde{N}\lambda} = \det(\Lambda)^{-1}\kappa_2 h^2 \Omega(\mathbf{u}_0)\sum_{t=1}^{d} a_t^{-2} \beta''_{tt}(\mathbf{u}_0) + o(h^2), \text{ if and only if } \lambda=1 \quad (3\text{-}2)$$

$$\beta_{ss}^{(2)}(\mathbf{u}_0) = \left( \left.\frac{\partial^2 \beta_1}{\partial u_s^2}\right|_{\mathbf{u}_0}, \ldots, \left.\frac{\partial^2 \beta_p}{\partial u_s^2}\right|_{\mathbf{u}_0} \right)^\mathrm{T} \text{ and } \Pi(\mathbf{U},\mathbf{u}_0) = ((\mathbf{U}-\mathbf{u}_0)\mathcal{H}\beta_k(\mathbf{u}_0)(\mathbf{U}-\mathbf{u}_0)^\mathrm{T})_{k=1,\ldots,p}^\mathrm{T},$$

where $\mathcal{H}\beta_k(\mathbf{u}_0)$ denotes the Hessian matrix of function $\beta_k(\mathbf{u})$ when $\mathbf{u} = \mathbf{u}_0$.

**Lemma 3**. Under the same preconditions as lemma 1, while $l=d, d+1$ and $d-1$. we have these results, set $C_{\widetilde{N}\lambda}^s = (\widetilde{N}h^{2l})^{-1}\sum_{\mathbf{i}\in\mathbf{I}} K_h(\frac{d_0(\mathbf{U_i})}{h})(\mathbf{U_{is}}-u_{0s})^\lambda \sigma(\mathbf{U_i}) X_\mathbf{i}^\mathrm{T} X_\mathbf{i}$, then

$$C_{\widetilde{N}\lambda}^s = \det(\Lambda) a_s^\lambda)^{-1} h^{(\lambda+d)-(2l+1)} \left( \int v_s^\lambda \left(\prod_{r=1}^d K(v_r)\right)^2 dv \right) \sigma(\mathbf{u}_0) f(\mathbf{u}_0) \Omega(\mathbf{u}_0) + o(h^{(\lambda+d)-(2l+1)}). \quad (3\text{-}3)$$

(See proof of lemmas in appendix.) Therefore, based on the preliminary work above, one of the major results of this paper is obtained as follow,

**Theorem 1**. Under assumptions from A1 to A6 and lemma 1,2 and 3, when $\mathbf{u}_0$ is an interior point of $S(\mathbf{U})$, we have the following results,

$$Bias(\hat{\beta}(\mathbf{u}_0) \mid X, \mathbf{U}) = \frac{\kappa_2 h^2}{2} \Omega^{-1}(\mathbf{u}_0) \sum_{s=1}^d a_s^{-2} \beta_{ss}^{(2)}(\mathbf{u}_0) + o(h^2), \quad (3\text{-}5)$$

$$Var(\hat{\beta}(\mathbf{u}_0) \mid X, \mathbf{U}) = (\det(\Lambda)\widetilde{N}h^{(d-1)})^{-1} \varphi(\mathbf{u}_0) \kappa \sigma(\mathbf{u}_0) f(\mathbf{u}_0) \Omega(\mathbf{u}_0)\{1+o_p(1)\}, \quad (3\text{-}6)$$

where $\kappa = \int \prod_{r=1}^d K^2(v_r) dv$, $\kappa_2 = \int z^2 K(z) dz$, $v = (v_1, v_2, \ldots, v_d)$, $z \in \mathbf{R}$. $\varphi(\mathbf{u})$ is a $p \times p$ matrix purely based on the value of vector $\mathbf{u}$ and it can be rewritten as this, $\varphi(\mathbf{u}) = \mathbf{Q}_{11}\mathbf{Q}_{11}^\mathrm{T}$, where $\mathbf{Q}_{11}$ is the upper-left block of matrix $\mathbf{Q}$.

(3-5) and (3-6) clearly demonstrate that the value of scale parameters will only affect conditional bias. Note that conditional variances of the estimation of each covariate are exactly the diagonal entries of conditional variance matrix. On the other hand, due to the construction of (2-2), it can be rewritten as follow,

$$\hat{\beta}(\mathbf{u}_0) = e_1^\mathrm{T} \left( \widetilde{N}^{-1} \sum_{\mathbf{i}\in\mathbf{I}} K_h(\tfrac{d_0(\mathbf{u_i})}{h}) \widetilde{X}_\mathbf{i}^\mathrm{T} \widetilde{X}_\mathbf{i} \right)^{-1} \left( \widetilde{N}^{-1} \sum_{\mathbf{i}\in\mathbf{I}} K_h(\tfrac{d_0(\mathbf{u_i})}{h}) \widetilde{X}_\mathbf{i}^\mathrm{T} y_\mathbf{i} \right). \quad (3\text{-}7)$$

Note that, by using the technique in appendix, matrix $\widetilde{N}^{-1}\sum_{\mathbf{i}\in\mathbf{I}} K_h(\tfrac{d_0(\mathbf{u_i})}{h}) \widetilde{X}_\mathbf{i}^\mathrm{T} \widetilde{X}_\mathbf{i}$ is asymptotically equal to a zero matrix which will bring us much difficulty of analysis. To overcome this unnecessary barrier, based on the property of inverse matrix, the following equivalent transformation is proposed,

set $$G = \begin{bmatrix} h^{-(d-1)}I_{p\times p} & \mathbf{0} \\ \mathbf{0} & h^{-(d+1)}I_{dp\times dp} \end{bmatrix},$$

where $I_{p\times p}$ and $I_{dp\times dp}$ denote $p\times p$ and $dp\times dp$ identity matrix respectively, then we have

$$\hat{\beta}(\mathbf{u}_0) = e_1^{\mathrm{T}}\left(\widetilde{N}^{-1}\sum_{\mathbf{i}\in\mathbf{I}}K_h(\tfrac{d_0(\mathbf{u_i})}{h})\widetilde{X}_{\mathbf{i}}^{\mathrm{T}}\widetilde{X}_{\mathbf{i}}\right)^{-1}G^{-1}G\left(\widetilde{N}^{-1}\sum_{\mathbf{i}\in\mathbf{I}}K_h(\tfrac{d_0(\mathbf{u_i})}{h})\widetilde{X}_{\mathbf{i}}^{\mathrm{T}}y_{\mathbf{i}}\right), \quad (3\text{-}8)$$

$$= e_1^{\mathrm{T}}\left(\widetilde{N}^{-1}\sum_{\mathbf{i}\in\mathbf{I}}K_h(\tfrac{d_0(\mathbf{u_i})}{h})G\widetilde{X}_{\mathbf{i}}^{\mathrm{T}}\widetilde{X}_{\mathbf{i}}\right)^{-1}\left(\widetilde{N}^{-1}\sum_{\mathbf{i}\in\mathbf{I}}K_h(\tfrac{d_0(\mathbf{u_i})}{h})G\widetilde{X}_{\mathbf{i}}^{\mathrm{T}}y_{\mathbf{i}}\right), \quad (3\text{-}9)$$

or based on (2-3), we have

$$\hat{\beta}(\mathbf{u}_0) = e_1^{\mathrm{T}}\left(G\widetilde{X}^{\mathrm{T}}W(\mathbf{u}_0)\widetilde{X}\right)^{-1}\left(G\widetilde{X}^{\mathrm{T}}W(\mathbf{u}_0)y\right). \quad (3\text{-}10)$$

where $W(\mathbf{u}_0) = diag(K_h(\tfrac{d_0(\mathbf{u_1})}{h}),\ldots,K_h(\tfrac{d_0(\mathbf{u_N})}{h}))$ and $\mathbf{N} = (N_1,\ldots,N_M)$. Hence, the whole proof of theorem 1 is based on (3-9) and (3-10).

**Proof of Theorem 1**:

As for (3-5), due to (3-9), we firstly use Taylor series to expand each $y_{\mathbf{i}}$ as follow,

$$y_{\mathbf{i}} = \sum_{k=1}^{p} X_{\mathbf{i}k}\left(\beta_k(\mathbf{u}_0) + (\mathbf{u}_i - \mathbf{u}_0)\beta_k'(\mathbf{u}_0) + \tfrac{1}{2}(\mathbf{u} - \mathbf{u}_0)\mathcal{H}\beta_k(\mathbf{u}_0)(\mathbf{u} - \mathbf{u}_0)^{\mathrm{T}} + r\right) + \sigma(\mathbf{U_i})\varepsilon_{\mathbf{i}}$$

$$= \widetilde{X}_{\mathbf{i}}\mathbf{B}(\mathbf{u}_0) + \tfrac{1}{2}X_{\mathbf{i}}\Pi(\mathbf{u_i},\mathbf{u}_0) + \sigma(\mathbf{U_i})\varepsilon_{\mathbf{i}}, \quad (3\text{-}11)$$

where $\mathbf{B}(\mathbf{u}_0) = (\beta^{\mathrm{T}}(\mathbf{u}_0),\ \beta_1^{'\mathrm{T}}(\mathbf{u}_0),\ldots,\beta_d^{'\mathrm{T}}(\mathbf{u}_0))^{\mathrm{T}}$ and r is the remainder of Taylor expansion. Take (3-11) back to (3-9), we get

$$Bias(\hat{\beta}(\mathbf{u}_0)\mid X,\mathbf{U}) = e_1^{\mathrm{T}}\left(\widetilde{N}^{-1}\sum_{\mathbf{i}\in\mathbf{I}}K_h(\tfrac{d_0(\mathbf{u_i})}{h})G\widetilde{X}_{\mathbf{i}}^{\mathrm{T}}\widetilde{X}_{\mathbf{i}}\right)^{-1}\left(\widetilde{N}^{-1}\sum_{\mathbf{i}\in\mathbf{I}}K_h(\tfrac{d_0(\mathbf{u_i})}{h})G\widetilde{X}_{\mathbf{i}}^{\mathrm{T}}X_{\mathbf{i}}\Pi(\mathbf{u_i},\mathbf{u}_0)\right)$$

Note the term of remainder is of negligible order compare to the term arising from the equation above, thus we skip it here. Define $\mathbf{A}^{1,\mathbf{u}} = \widetilde{N}^{-1}\sum_{\mathbf{i}\in\mathbf{I}}K_h(\tfrac{d_0(\mathbf{u_i})}{h})G\widetilde{X}_{\mathbf{i}}^{\mathrm{T}}\widetilde{X}_{\mathbf{i}}$. This matrix can be partitioned as follow,

$$\widetilde{N}^{-1}\sum_{\mathbf{i}\in\mathbf{I}}K_h(\tfrac{d_0(\mathbf{u_i})}{h})\begin{bmatrix} h^{-(d-1)}X_{\mathbf{i}}^{\mathrm{T}}X_{\mathbf{i}} & h^{-(d-1)}(\mathbf{u_i}-\mathbf{u}_0)\otimes X_{\mathbf{i}} \\ h^{-(d+1)}(\mathbf{u_i}-\mathbf{u}_0)^{\mathrm{T}}\otimes X_{\mathbf{i}}^{\mathrm{T}} & h^{-(d+1)}((\mathbf{u_i}-\mathbf{u}_0)\otimes X_{\mathbf{i}})^{\mathrm{T}}(\mathbf{u_i}-\mathbf{u}_0)\otimes X_{\mathbf{i}} \end{bmatrix}. \quad (3\text{-}12)$$

and applying the results from lemma 1, the following results can be shown,

$$(\tilde{N}h^{(d-1)})^{-1}\sum_{\mathbf{i}\in\mathbf{I}}K_h(\tfrac{d_0(\mathbf{u_i})}{h})X_\mathbf{i}^\mathrm{T}X_\mathbf{i} == \det(\Lambda)^{-1}\underbrace{\Omega(\mathbf{u}_0)f(\mathbf{u}_0)}_{\mathbf{A}_{11}^{1,\mathbf{u}}}+o_p(1) \quad (3\text{-}13)$$

$$(\tilde{N}h^{(d-1)})^{-1}\sum_{\mathbf{i}\in\mathbf{I}}K_h(\tfrac{d_0(\mathbf{u_i})}{h})(\mathbf{u_i}-\mathbf{u}_0)\otimes X_\mathbf{i}$$
$$=\det(\Lambda)^{-1}\underbrace{h^2\kappa_2\left[(\Lambda^{-2}\otimes I_{d\times d})(\Gamma'(\mathbf{u}_0)f(\mathbf{u}_0)+f'(\mathbf{u}_0)\otimes\Gamma(\mathbf{u}_0))\right]^\mathrm{T}}_{\mathbf{A}_{12}^{1,\mathbf{u}}}\{1+o_p(1)\}, \quad (3\text{-}14)$$

$$(\tilde{N}h^{(d+1)})^{-1}\sum_{\mathbf{i}\in\mathbf{I}}K_h(\tfrac{d_0(\mathbf{u_i})}{h})(\mathbf{u_i}-\mathbf{u}_0)\otimes X_\mathbf{i}$$
$$=\det(\Lambda)^{-1}\underbrace{\kappa_2\left[(\Lambda^{-2}\otimes I_{d\times d})(\Gamma'(\mathbf{u}_0)f(\mathbf{u}_0)+f'(\mathbf{u}_0)\otimes\Gamma(\mathbf{u}_0))\right]^\mathrm{T}}_{\mathbf{A}_{21}^{1,\mathbf{u}}}+o_p(1), \quad (3\text{-}15)$$

$$(\tilde{N}h^{d+1})^{-1}\sum_{\mathbf{i}\in\mathbf{I}}K_h(\tfrac{d_0(\mathbf{u_i})}{h})[(\mathbf{u_i}-\mathbf{u}_0)\otimes X_\mathbf{i}]^\mathrm{T}[(\mathbf{u_i}-\mathbf{u}_0)\otimes X_\mathbf{i}]$$
$$=\det(\Lambda)^{-1}\underbrace{\Lambda^{-2}(\kappa_2\Omega(\mathbf{u}_0)f(\mathbf{u}_0))I_{dp\times dp}}_{\mathbf{A}_{22}^{1,\mathbf{u}}} \quad (3\text{-}16)$$

Based on (3-13) to (3-16), we can tell the inverse matrix of $\mathbf{A}^{1,\mathbf{u}}=\det(\Lambda)^{-1}\begin{bmatrix}\mathbf{A}_{11}^{1,\mathbf{u}} & \mathbf{A}_{12}^{1,\mathbf{u}} \\ (\mathbf{A}_{21}^{1,\mathbf{u}})^\mathrm{T} & \mathbf{A}_{22}^{1,\mathbf{u}}\end{bmatrix}$ is asymptotically existed, denoted by $\mathbf{Q}$.

Define $\mathbf{A}^{2,\mathbf{u}} = \tilde{N}^{-1}\sum_{\mathbf{i}\in\mathbf{I}}K_h(\tfrac{d_0(\mathbf{u_i})}{h})G\tilde{X}_\mathbf{i}^\mathrm{T}X_\mathbf{i}\Pi(\mathbf{u_i},\mathbf{u}_0)$ which can be partitioned as follow,

$$\mathbf{A}^{2,\mathbf{u}} = \tilde{N}^{-1}\sum_{\mathbf{i}\in\mathbf{I}}K_h(\tfrac{d_0(\mathbf{u_i})}{h})\begin{bmatrix} h^{-(d-1)}X_\mathbf{i}^\mathrm{T}X_\mathbf{i}\Pi(\mathbf{u_i},\mathbf{u}_0) \\ h^{-(d+1)}(\mathbf{u_i}-\mathbf{u}_0)^\mathrm{T}X_\mathbf{i}^\mathrm{T}X_\mathbf{i}\Pi(\mathbf{u_i},\mathbf{u}_0) \end{bmatrix}.$$

Similar to the calculation of $\mathbf{A}^{1,\mathbf{u}}$, we can simply obtain that

$$\mathbf{A}^{2,\mathbf{u}} = \begin{bmatrix} \tfrac{1}{2}\det(\Lambda)^{-1}\kappa_2 h^2 f(\mathbf{u}_0)\Omega(\mathbf{u}_0)\sum_{s=1}^d a_s^{-2}\beta_{ss}^{(2)}(\mathbf{u}_0)+o(h^2) \\ \mathbf{0} \end{bmatrix}.$$

Finally, to construct $\mathbf{Q}$, we need the following tricks(see, for example Li and Racine(2015), p82,

Chapter 2) . Set $\Xi = (\mathbf{A}_{22}^{1,\mathbf{u}} - (\mathbf{A}_{21}^{1,\mathbf{u}})^\mathrm{T}(\mathbf{A}_{11}^{1,\mathbf{u}})^{-1}\mathbf{A}_{12}^{1,\mathbf{u}})^{-1}$, then we have

$$\mathbf{Q} = \det(\Lambda)\begin{pmatrix} (\mathbf{A}_{11}^{1,\mathbf{u}})^{-1}(I_{p\times p}+\mathbf{A}_{12}^{1,\mathbf{u}}\Xi(\mathbf{A}_{21}^{1,\mathbf{u}})^\mathrm{T}(\mathbf{A}_{11}^{1,\mathbf{u}})^{-1}) & -(\mathbf{A}_{11}^{1,\mathbf{u}})^{-1}\mathbf{A}_{12}^{1,\mathbf{u}}\Xi \\ -\Xi(\mathbf{A}_{21}^{1,\mathbf{u}})^\mathrm{T}(\mathbf{A}_{11}^{1,\mathbf{u}})^{-1} & \Xi \end{pmatrix}. \quad (3\text{-}17)$$

By using some simple knowledge of matrix algebra, we can finish the proof of (3-5). As for (3-6), based on (3-10), conditional variance matrix $Var(\hat{\beta}(\mathbf{u}_0)|X,\mathbf{U})$ can be firstly expressed as follow,

$$e_1^T(G\widetilde{X}^TW(\mathbf{u}_0)\widetilde{X})^{-1}(G\widetilde{X}^TW(\mathbf{u}_0)VW(\mathbf{u}_0)\widetilde{X}G)(\widetilde{X}^TW(\mathbf{u}_0)\widetilde{X}G)^{-1}e_1, \qquad (3\text{-}18)$$

where $V = diag(\sigma(\mathbf{u}_1),\ldots,\sigma(\mathbf{u}_N))$. Set $\Psi = G\widetilde{X}^TW(\mathbf{u}_0)VW(\mathbf{u}_0)\widetilde{X}G$, (3-17) is equal to

$$e_1^T(\widetilde{N}^{-1}G\widetilde{X}^TW(\mathbf{u}_0)\widetilde{X})^{-1}\widetilde{N}^{-1}(\widetilde{N}^{-1}\Psi)(\widetilde{N}^{-1}\widetilde{X}^TW(\mathbf{u}_0)\widetilde{X}G)^{-1}e_1, \qquad (3\text{-}19)$$

and by using lemma 3, we directly obtain

$$\widetilde{N}^{-1}\Psi = \det(\Lambda)^{-1}\begin{bmatrix} h^{-(d-1)}\kappa f(\mathbf{u}_0)\Omega(\mathbf{u}_0) & h^{-d}\left(\int v\Lambda^{-1}\prod_{r=1}^d K^2(v_r)dv\right)\sigma(\mathbf{u}_0)f(\mathbf{u}_0)\Omega(\mathbf{u}_0) \\ * & h^{-(d+2)}\left(\int \Lambda^{-1}v^Tv\Lambda^{-1}\prod_{r=1}^d K^2(v_r)dv\right)\sigma(\mathbf{u}_0)f(\mathbf{u}_0)\Omega(\mathbf{u}_0) \end{bmatrix} + o_p(1)$$

Again, by using (3-17) and some simple knowledge of matrix product, we obtain (3-6). Due to (3-5) and (3-6), it can be easily figured out how the value of scale parameters affect the leading terms of conditional bias and variance, which also demonstrate the mean-square consistency of **GWLE**.

***Corollary 1***. *Under assumptions A1 to A6 and lemma 1,2,3, for any given interior point* $\mathbf{u}_0$, *the following results can be obtained,*

$$\hat{\beta}(\mathbf{u}_0) - \beta(\mathbf{u}_0) = O_p(dh^2 + (\widetilde{N}h^{(d-1)})^{-\frac{1}{2}}), \qquad (3\text{-}20)$$

$$\lim_{\widetilde{N}\to+\infty}\frac{Var(\hat{\beta}_{GWLE}(\mathbf{u}_0))}{Var(\hat{\beta}_{MWLE}(\mathbf{u}_0))} = 0. \qquad (3\text{-}21)$$

Proof of this corollary is obvious. We can directly obtain (3-20) by (3-5) and (3-6). As for (3-21), by following the same technique demonstrated(or see, for example Ruppert and Wand(1994)) in the proof of theorem 1 and lemmas, we could obtain $Var(\hat{\beta}_{MWLE}(\mathbf{u}_0)) = O(\widetilde{N}|\mathbf{H}|)^{-1}$, where $\mathbf{H}$ is the bandwith matrix and $|.|$ denotes the determinant. Without loss of generality, by assuming $\mathbf{H}$ is a symmetric and positive definite matrix, we can obtain (3-21). More specifically, when the multidimensional kernel is a product kernel, $Var(\hat{\beta}_{MWLE}(\mathbf{u}_0)) = O(\widetilde{N}\prod_{s=1}^d h_s)^{-1}$.

Together with all the discussion above, we clearly prove that **GWLE** is asymptotically more efficient than **MWLE**.

Our second task is to investigate this question: How do the scale parameters affect the selection of globally optimal bandwith. Similar to the technique we use to discover golden bandwith, we here still choose to minimize the mean integrated squared error at first. Please recall that our estimator $\hat{\beta}(\mathbf{u}_0)$ is a vector. Though variance of each entry of $\hat{\beta}(\mathbf{u}_0)$ has been contained by its variance matrix, in this paper, we prefer to a more asymptotic view. That is to say, calculation of some constant coefficient purely based on $\mathbf{u}_0$ will not be taken into consideration. Therefore, we obtain the corollary 2.

***Corollary 2.*** *Under the same preconditions as **Corollary 1**, based on **Theorem 1**, we obtain the following relation among golden bandwith, sample size and scale parameters,*

$$h_{opt} = \left( \frac{2\widetilde{N}\det(\Lambda)}{d} \sum_{s=1}^{d} a_s^{-4} \right)^{-\frac{1}{d+2}} \left( \int C_s(\mathbf{u})^{-\frac{1}{d+2}} d\mathbf{u} \right),$$

*where* $C_s(\mathbf{u}) = \left( \Omega^{-1}(\mathbf{u}_0) \beta_{ss}^{(2)}(\mathbf{u}_0) \right) * \left( diag(\varphi(\mathbf{u}_0) \kappa \sigma(\mathbf{u}_0) f(\mathbf{u}_0) \Omega(\mathbf{u}_0))^{-1} \mathbf{1}_p \right)$. *For any* $n \times n$ *matrix* $\mathbf{S}$, $diag(\mathbf{S})$ *here indicates a diagonal matrix which only consists of diagonal entries of* $\mathbf{S}$.

Proof of this corollary is exactly the same as obtaining optimal bandwith, so we skip it here. It is clear that the optimal bandwith is $O(\widetilde{N}^{-\frac{1}{d+2}})$.

## 4. Conclusion

In this paper, we demonstrate the conditional bias and variance of locally linear GWR estimator for any coefficient function satisfying some smoothing conditions. Meanwhile, the relationship between the two properties and design of scale parameters is also revealed. Due to our theoretical results, it can been see that value of scale parameters could only affect the conditional bias of estimator.

## 5. Appendix

In this section, we demonstrate the proof of lemmas used before. However, considering proof of the three lemmas are essentially the same, we only prove lemma 1 here. The argument works the same to the other two.

**Proof of Lemma 1**,

Denote $K_h(\frac{d_0(\mathbf{U_i})}{h})h^{-l}(\mathbf{U}_{is}-u_{0s})^\lambda X_\mathbf{i}^\mathrm{T} X_\mathbf{i} = \delta_\mathbf{i}^s$ and $K_h(\frac{d_0(\mathbf{U})}{h})h^{-l}(\mathbf{U}_s-u_{0s})^\lambda X^\mathrm{T} X = \delta^s$, it is obvious that, under the assumption of stationarity, we have $E(A_{\tilde{N}\lambda}^s) = E(\delta^s)$. By using law of iterated expectation,

$$E\big(E(\delta^s \mid \mathbf{U}=\mathbf{u})\big) = h^{-(l+1)}\int K(\frac{d_0(\mathbf{u})}{h})(u_s-u_{0s})^\lambda \Omega(\mathbf{u})f(\mathbf{u})d\mathbf{u} \qquad (A\text{-}1)$$

$\because K(\frac{d_0(\mathbf{u})}{h}) = \phi((\frac{d_0(\mathbf{u})}{h})^2) = \phi(\sum_{r=1}^d (\frac{a_r(u_r-u_{0r})}{h})^2) = \prod_{r=1}^n \phi((\frac{a_r(u_r-u_{0r})}{h})^2)$,

set $v_r = \frac{a_r(u_r-u_{0r})}{h}$, then $\mathbf{u} = \mathbf{u}_0 + v\Lambda^{-1}h$ and the Jacobian determinant is $\det(\Lambda^{-1})h$, hence

$$(A\text{-}1) = (\det(\Lambda)a_s^\lambda)^{-1} h^{(\lambda+d)-(l+1)} \int v_s^\lambda \prod_{r=1}^n K(v_r)\Omega(\mathbf{u}_0+v\Lambda^{-1}h)f(\mathbf{u}_0+v\Lambda^{-1}h)dv \,.(A\text{-}2)$$

By using Taylor expansion, we have

$$\Omega(\mathbf{u}_0+v\Lambda^{-1}h) = \Omega(\mathbf{u}_0) + h\sum_{t_1=1}^d \Omega'_{t_1}(\mathbf{u}_0)(a_{t_1}^{-1}v_{t_1}) + o(h), \qquad (A\text{-}3)$$

$$f(\mathbf{u}_0+v\Lambda^{-1}h) = f(\mathbf{u}_0) + h\sum_{t_2=1}^d f'_{t_2}(\mathbf{u}_0)(a_{t_2}^{-1}v_{t_2}) + o(h). \qquad (A\text{-}4)$$

Take (A-3) and (A-4) back to (A-2), we instantly get

$$\begin{aligned}(A\text{-}1) = (\det(\Lambda)a_s^\lambda)^{-1} h^{(\lambda+d)-(l+1)} \{&\kappa_\lambda \Omega(\mathbf{u}_0)f(\mathbf{u}_0) \\ &+ h\kappa_{\lambda+1}[\Omega'_s(\mathbf{u}_0)f(\mathbf{u}_0)+\Omega(\mathbf{u}_0)f'_s(\mathbf{u}_0)]\} + o(1)\end{aligned} \qquad (A\text{-}5)$$

Note that $A_{\tilde{N}\lambda}^s$ is a matrix. In order to investigate the convergence of every entries of $A_{\tilde{N}\lambda}^s$, we define for any $p\times q$ matrix $\mathbf{W}_1$ and $\mathbf{W}_2$,

$$HVar(\mathbf{W}_1) = E(\mathbf{W}_1 * \mathbf{W}_1) - E(\mathbf{W}_1) * E(\mathbf{W}_1),$$

$$HCov(\mathbf{W}_1, \mathbf{W}_2) = E(\mathbf{W}_1 * \mathbf{W}_2) - E(\mathbf{W}_1) * E(\mathbf{W}_2),$$

which actually represent the variance of every entry and covariance of every two entries from $\mathbf{W}_1$ and $\mathbf{W}_2$ respectively having the same position. Therefore,

$$HVar(A^s_{\tilde{N}\lambda}) = \tilde{N}^{-2} HVar(\sum_{\mathbf{i} \in \mathbf{I}} \delta^s_\mathbf{i}) = \tilde{N}^{-2} \{\sum_{\mathbf{i} \in \mathbf{I}} HVar(\delta^s_\mathbf{i}) + \sum_{\mathbf{i} \neq \mathbf{j}} HCov(\delta^s_\mathbf{i}, \delta^s_\mathbf{j})\}.$$

Due to assumption **A**1, the following inequality can be directly obtained,

$$\sum_{\mathbf{i} \neq \mathbf{j}} HCov(\delta^s_\mathbf{i}, \delta^s_\mathbf{j}) = \sum_\mathbf{i} \sum_{\|\mathbf{j}-\mathbf{i}\|_\infty \leq m} HCov(\delta^s_\mathbf{i}, \delta^s_\mathbf{j}) \leq \sum_\mathbf{i} \sum_{\|\mathbf{j}-\mathbf{i}\|_\infty \leq m} \rho(m) HVar(\delta^s_\mathbf{i})$$

$$\leq (m^M - 1)\rho(m) \sum_\mathbf{i} HVar(\delta^s_\mathbf{i}) = \tilde{N}(m^M - 1)\rho(m) HVar(\delta^s)$$

Take it back, owing to stationarity, we get

$$HVar(A^s_{\tilde{N}\lambda}) \leq \tilde{N}^{-1}(m^M - 1)\rho(m) + 1) HVar(\delta^s),$$

where $m$, $M$ and $\rho(m)$ are all fixed constants. Furthermore, similar to the technique used before, the following result can be obtained,

$$E(\delta^s * \delta^s) = (\det(\Lambda) a_s^{2\lambda})^{-1} h^{2(\lambda-l-1)+d} \left(\int v_s^{2\lambda} \prod_{r=1}^n K(v_r) dv\right) \Omega(\mathbf{u}_0) f(\mathbf{u}_0)\{1 + o_p(1)\}.$$

Together with (A-5) and assumption A6, we obtain

$$HVar(A^s_{\tilde{N}\lambda}) = o(1).$$